# DIRAC'S REDUCED RADI AL EQUATIONS AND THE PROBLEM OF ADDITIONAL SOLUTIONS

ANZOR KHELASHVILI[1,2,], TEIMURAZ NADAREISHVILI[1,3,a]

[1] *Inst. of High Energy Physics, Iv. Javakhishvili Tbilisi State University, University Str. 9, 0109, Tbilisi, Georgia. E-mail:anzor.khelashvili@tsu.ge*

[2] *St.Andrea the First-called Georgian University of Patriarchate of Georgia, Chavchavadze Ave.53a, 0162, Tbilisi,Georgia.*

[3] *Faculty of Exact and Natural Sciences, Iv. Javakhishvili Tbilisi State University, Chavchavadze Ave 3, 0179, Tbilisi,Georgia.*

Corresponding author   E-mail: *teimuraz.nadareishvili@tsu.ge*

We show that additional solutions must be ignored (in differences of the Schrodinger and Klein-Gordon equations) in the Dirac equation, where usually passed the second order radial equation, called the reduced equation, instead of a system. Analogously to the Schrodinger equation, in this process the Dirac's delta function appears, which was unnoted during the full history of quantum mechanics. This unphysical term we remove by a boundary condition at the origin. However, the distribution theory imposes on the radial function strong restriction and by this reason practically for all potentials, even regular, use of these reduced equations is not permissible. At the end we include consideration in the framework of two-dimensional Dirac equation. We show that even here the additional solution does not survives as a result of usual physical requirements.

## 1. Introduction

Time by time there appear articles [1-6], in which authors considered so-called additional solutions in problem of hydrogen-like atoms. These solutions are related to the $1/r^2$

---

[a] *Corresponding author*

Anzor Khelashvili, Teimuraz Nadareishvili

behaving term in potential of Schrodinger or other related equations. They are called by various names: additional, peculiar levels, sometimes – as a small hydrogen ("hydrino'). It is remarkable to note, that some authors relate these solutions to the different physical phenomena, such as a specific radiation of Galaxy, dark matter [6] or the cold fusion [7,8] in nuclear reactions, etc.

In the earlier papers [9-12] was shown that this kind of solutions really exist in the Klein-Gordon equation for Coulomb potential and their status is established after applying of the self-adjoint extension procedure (SAE). Some authors believed that we must have exactly the same situation in the Dirac equation, so-called Dirac's deep levels (DDL) exists [13]. Contrary to them, we consider below this problem more carefully in the framework of the Dirac radial equation. It turns out that the fulfillment to all general principles of quantum mechanics for additional solution is impossible and the final conclusion is, probably, pessimistic.

Our approach is based on the radial Dirac equations for Coulomb potential. While the results following from the Dirac equation often are well-known, we'll try below to get exhaustive analysis.

The paper is constructed in the following manner: First of all we reexamine the normalization condition of the Dirac radial functions and their corresponding behavior at the origin of coordinates. We show that existence of additional states is problematic. Then we pass to unique equations for reduced wave functions, which consist only second derivative. There appears an extra term, consisting a delta function, which has no physical sense. To avoid this term we must impose boundary condition-like constraint. Moreover, from the distribution theory ground this constraint further restricts radial functions by power-like asymptotes at the origin with integer degrees. We investigate their behavior at the origin for various potentials. We show that no ones satisfy to strong mathematical (in viewpoint of theory of distributions) requirements even for regular solutions. At the Appendix B we consider the same problem in the framework of two dimensional Dirac equation. We show that the additional solutions still absent.

## 2. Dirac's radial equations and Normalization properties of Radial equations

As is well-known, to derive radial equations the full Dirac wave function is represented in the following form [14]

$$\psi(\vec{r}) = \begin{pmatrix} f(r)\Omega_{jlm}(\vec{n}) \\ \left((-1)^{\frac{1+l-l'}{2}} g(r)\Omega_{jl'm}(\vec{n})\right) \end{pmatrix}; \quad l = j \pm 1/2, \; l' = 2j - l \qquad (1)$$

Here $\Omega_{jlm}$ are spin-spherical harmonics, depended on spherical angles $\hat{n}$. Substituting Eq. (1) into the Dirac equation gives the system of first order radial equations [14]

$$f' + \frac{1+\kappa}{r}f - (E+m+S-V)g = 0$$
$$g' + \frac{1-\kappa}{r}g + (E-m-S-V)f = 0$$
(2)

Where together with vector potential $V(r)$ a scalar potential $S(r)$ is also included. Here

$$\kappa = \begin{cases} -(j+1/2) = -(l+1), & \text{if } j = l+1/2 \\ (j+1/2) = l, & \text{if } j = l+1/2 \end{cases}$$
(3)

is the Dirac quantum number, which corresponds to the conserved Dirac matrix in any central potential

$$K = \beta(\vec{\Sigma}\vec{l} + 1)$$
(4)

In many applications the system (2) is rewritten for combinations

$$F(r) = rf(r) \quad and \quad G(r) = rg(r)$$
(5)

Then it takes the form

$$F' + \frac{\kappa}{r}F - (E+m+S-V)G = 0$$
$$G' - \frac{\kappa}{r}G + (E-m-S-V)F = 0$$
(6)

Now the problem left is: what kind of asymptotic follows from the equations (2) or (6) for radial functions? It is evident that according to (2) or (6) the same asymptotics for both functions follow only for Coulomb potential.

In general, the behavior of radial wave function at the origin of coordinates is closely related to the normalization property of wave function. As in the Schrodinger case this behavior depends on what are required to be finite – probability density, full probability or differential probability [11].

For example, finiteness of differential probability in the spherical slice $(r, r+dr)$

$$|f|^2 r^2 dr < \infty, \quad |g|^2 r^2 dr < \infty$$
(7)

gives $f, g \approx r^\lambda$ at the origin, with $\lambda > -1$. On the other hand, finiteness of total probability inside a sphere of small radius $a$,

$$\int_0^a |f|^2 r^2 dr < \infty, \quad \int_0^a |g|^2 r^2 dr < \infty$$
(8)

permits for more singularity $\lambda > -3/2$. The same behavior follows also from finiteness of the norm.

Anzor Khelashvili, Teimuraz Nadareishvili

In order not to repeat well-known procedure [11], below we consider only the most general and strong restriction after W. Pauli [15] –time-independence of the total probability

$$\frac{d}{dt}\int \psi^+ \psi dV = 0 \tag{9}$$

Let us use the time-dependent Dirac equation

$$i\frac{\partial \psi}{\partial t} = H\psi, \quad H = \vec{\alpha}\cdot\vec{p} + \beta m + V(r) \tag{10}$$

Then the condition (9) takes the form

$$-i\int\left[\psi^+(H\psi) - (H\psi^+)\psi\right]dV = 0 \tag{11}$$

In other words, time-independence of probability means, that the Hamiltonian must be a self-adjoint operator.

Let us begin from the continuity equation in the Dirac problem

$$\frac{\partial \rho}{\partial t} + div\vec{j} = 0 \tag{12}$$

Where

$$\rho = \psi^+\psi, \quad \vec{j} = \psi^+\vec{\alpha}\psi \tag{13}$$

According to the continuity equation, conservation of the norm takes the following form after using the Gauss theorem

$$\frac{\partial}{\partial t}\int_V \psi^+\psi dV = -\int_V div\vec{j}\, dV = -\int_S j_N dS \tag{14}$$

where $j_N$ is the normal component of current into the surface.

If the Hamiltonian has a singular point at $r = 0$, caused by a potential, the Gauss' theorem in Eq.(14) is not applicable. We must exclude this point from the integration volume and surround it by a small sphere of radius $a$. In this case the surface integral is divided into a surface in infinity that encloses the total volume, and the surface of a sphere of radius $a$:

$$\lim_{a\to 0} a^2 \int j_a d\Omega + \int_a j_N dS = 0 \tag{15}$$

In the first integral here we have expressed the surface element of the sphere as $dS = a^2 d\Omega$, where $d\Omega$ is an element of solid angle. Because the wave function must vanish at infinity (bound state), the second term goes to zero. As regards of normal component of $j_N$ (its radial part) using Eq.(1) and taking into account that

$$\alpha_r = \begin{pmatrix} 0 & -i \\ i & 0 \end{pmatrix} \tag{16}$$

it takes the form

Dirac's radial equations and the Additional Solutions

$$j_N = i(fg^\bullet - gf^\bullet)_{r=a} \tag{17}$$

Because $r$ is confined in a small region one can replace the radial functions by their behavior

$$\lim_{r \to 0} F(r) = \lim_{r \to 0} rf = 0 \tag{18}$$

$$\lim_{r \to 0} G(r) = \lim_{r \to 0} rg = 0 \tag{19}$$

## 3. On the existence of additional solutions

We now investigate problem of existence of additional solutions. Consider directly the Coulomb potential (for simplicity only vector case will be considered)

$$V = -\frac{V_0}{r}, V_0 > 0 \tag{20}$$

In this case radial equations (2) become (at the origin)

$$f' + \frac{1+\kappa}{r} f - \frac{V_0}{r} g = 0$$
$$g' + \frac{1-\kappa}{r} g + \frac{V_0}{r} f = 0 \tag{21}$$

As both functions enter here at equal footing, they can be taken at $r \to 0$ in the same form

$$f = Ar^\lambda \quad \text{and} \quad g = Br^\lambda \tag{22}$$

After substitution into equations we obtain the following characteristic condition

$$(\lambda+1)^2 = \kappa^2 - V_0^2 \tag{23}$$

or

$$\lambda = -1 \pm P; \ P = \sqrt{\kappa^2 - V_0^2} \tag{24}$$

It means that the radial wave functions behave at the origin as

$$\lim_{r \to 0} f(r) = a_1 r^{-1+P} + b_1 r^{-1-P} \equiv f_{st} + f_{add}$$
$$\lim_{r \to 0} g(r) = a_2 r^{-1+P} + b_2 r^{-1-P} \equiv g_{st} + g_{add} \tag{25}$$

If we compare this behavior to the corresponding result for Schrodinger equation [11, 12], it appears that now the singularity at origin is grown by factor $r^{-1/2}$. This difference is easily to understand even on the dimensional grounds of view, because the energy expressions in both cases look like

$$E_{Sch} = \int d^3 r \psi_{Sch}^\bullet(\vec{r}) \left(\frac{p^2}{2m}\right) \psi_{Sch}(\vec{r}); \ E_{Dr} = \int d^3 r \psi_{Dr}^+(\vec{r})(\vec{\alpha}\vec{p} + \beta m) \psi_{Dr}(\vec{r}) \tag{26}$$

Anzor Khelashvili, Teimuraz Nadareishvili

We see that the scale dimensions for wave functions in these two cases differ exactly by ½ degree. This fact has important influence on physical picture – the additional solutions behave like

$$f_{add}(g_{add}) = b_1(b_2)r^{-1-P} \tag{27}$$

while for the Schrodinger equation we had [11,12]

$$f_{add}(g_{add}) = b_1(b_2)r^{-1/2-P} \tag{28}$$

This difference shows that in Dirac equation we are unable to remain additional solution (because they don't obey (19) condition), as opposed to Schrodinger or Klein-Gordon equations, where all quantum-mechanical requirements may be satisfied in the $0 < P < 1/2$ interval [11, 12]. In the Appendix A, we return again to this problem making the role of spin more transparent.

There is also one stronger criterion, related to the self-adjointness of Hamiltonian, namely orthogonality of additional solutions. The solutions, mentioned above, in difference from the Schrodinger or Klein-Gordon equations, do not satisfy orthogonality condition as well [5,9]

$$\lim_{r \to 0}\{f_q^\bullet g_{q'} - f_{q'}g_q^\bullet\} = 0, \quad q^2 = 2mE \quad or \quad q^2 = E^2 - m^2 \tag{29}$$

Thus, we can say that for the Dirac equation the real existence of additional ("hydrino") state is doubtful.

Let us study now if there are additional states in the Dirac equation for regular attractive potentials, which have less singularity than the Coulomb one at the origin. For simplicity, consider only attractive vector potentials

$$\lim_{r \to 0} V(r) \approx -\frac{V_0}{r^\alpha}; V_0 > 0 \tag{30}$$

$\alpha$ may be arbitrary number in the interval [0,1]. It follows

a) $\quad f \underset{r \to 0}{\approx} A_1 r^l + B_1 r^{-l-1-\alpha}; \quad g \underset{r \to 0}{\approx} C_1 r^{l+1-\alpha} + D_1 r^{-l-2}, \text{ when } j = l+1/2 \tag{31}$

b) $\quad f \underset{r \to 0}{\approx} A_2 r^{l-\alpha} + B_2 r^{-l-1}; \quad g \underset{r \to 0}{\approx} C_2 r^{l-1} + D_2 r^{-l-\alpha}, \text{ when } j = l-1/2 \tag{32}$

It is evident that in these expressions we must remain only the first terms (see, e.g. [14], problem from Sec.35)

$$f \approx r^l, g \approx r^{l'-\alpha}; \quad \text{if } j = l+1/2 \tag{33}$$

$$f \approx r^{l-\alpha}, g \approx r^{l'}; \quad \text{if } j = l-1/2 \tag{34}$$

where $l' = 2j - l$.

As regards of second terms – they must be ignored, because they do not satisfy condition (19). We conclude that even in these cases no additional solutions exist. For

Dirac's radial equations and the Additional Solutions

comparison, remember that in the Schrodinger equation regular potentials are defined as $\lim_{r \to 0} r^2 V(r) = 0$ and only $R \approx r^l$ term is retained.

## 4. Reduction to a single equation

For more details the reduction to the equations for single functions $f$ and $g$ is convenient, which is achieved by exclusion of one of the functions from the system (2). For example, if we determine from the first equation of (2) the function $g = \frac{1}{E+m+S-V}\left\{f' + \frac{1+\kappa}{r}f\right\}$ and substitute it into the second equation, we are passing to the second order equation for a single function

$$f'' + \frac{2}{r}f' + \frac{V'-S'}{E+m+S-V}f' + \frac{V'-S'}{E+m-V+S}\frac{1+\kappa}{r}f + \\ + \left[(E-V)^2 - (m+S)^2 - \frac{\kappa(\kappa+1)}{r^2}\right]f = 0 \qquad (35)$$

The equation for the second function $g$ looks analogously, it can be derived from Eq.(35) by change $\kappa \to -\kappa$ and $(m+S) \to -(m+S)$ and has the form

$$g'' + \frac{2}{r}g' + \frac{V'-S'}{E+m+S-V}g' + \frac{V'-S'}{E+m-V+S}\frac{1+\kappa}{r}g + \\ + \left[(E-V)^2 - (m+S)^2 - \frac{\kappa(\kappa+1)}{r^2}\right]g = 0 \qquad (35a)$$

The equations (35,35a) consist both first and second order derivative terms. In this respect, they are analogous to the Schrodinger radial equation for full function $R$. But in the last equation the first derivative usually is withdrawn by substitution, which in our case has the form

$$f(r) = \frac{F(r)}{r}; \qquad g(r) = \frac{G(r)}{r} \qquad (36)$$

Remembering that the first two terms in Eqs. (35) are the radial parts of the Laplace operator and therefore

$$\left(\frac{d^2}{dr^2} + \frac{2}{r}\frac{d}{dr}\right)\left(\frac{1}{r}\right) = -4\pi\delta^{(3)}(\vec{r}) \qquad (37)$$

in the process of transition to the new functions (36) the additional delta function appears [11,16]. After using the spherical representation of 3-dimensional delta function and some trivial manipulations, we derive the following equation

$$rF'' - \delta(r)F(r) + \frac{V' - S'}{E + m + S - V}(rF' + \kappa F) + \\ + \left[(E - V)^2 - (m + S)^2 - \frac{\kappa(\kappa + 1)}{r^2}\right]rF = 0 \tag{38}$$

The term, containing delta-function has no physical sense and it must be avoided by boundary condition

$$F(0) = 0 \tag{39}$$

Moreover, in order that the product $\delta(r)F(r)$ be a well-defined distribution, the ordinary function $F(r)$ has to be infinitely smooth (infinitely differentiable) at $r = 0$. This requirement restricts $F(r)$ to be a power like function at the origin with integer degree

$$\lim_{r \to 0} F(r) \approx r^N; N = 0,1,2,... \tag{40}$$

It happens so because $r^\lambda \delta(r)$ product is not correctly defined as a distribution when $\lambda$ is not positive integer [17]

Taking into account all above mentioned, we are faced with the following radial equation for the reduced function $F(r)$

$$\left(\frac{d^2}{dr^2} - \frac{\kappa(\kappa + 1)}{r^2}\right)F(r) + \frac{V' - S'}{E + m + S - V}F' + \frac{\kappa}{r}\frac{(V' - S')}{E + m + S - V}F + \\ + \left[(E - V)^2 - (m + S)^2\right]F(r) = 0 \tag{41}$$

Note that the first derivative term still presents here and this equation has not final reduced form. One can eliminate also the remaining first derivative term from this equation by substitution

$$F(r) = \sqrt{E + m + S - V}\,\varphi(r) \tag{42}$$

and derive an equation for this new function $\varphi(r)$. It looks like

$$\frac{d^2\varphi(r)}{dr^2} + \left\{(E - V)^2 - (m + S)^2 - \frac{\kappa(\kappa + 1)}{r^2}\right\}\varphi(r) = \\ = \left\{\frac{3}{4}\frac{(V' - S')^2}{(E - V + m + S)^2} + \frac{V'' - S'' - \frac{2\kappa}{r}(V' - S')}{2(E - V + m + S)}\right\}\varphi(r) \tag{43}$$

The same arguments allow us to write down the equation for the second function

$$G(r) = \sqrt{E - m - S - V}\, \chi(r) \tag{44}$$

It has a form

$$\frac{d^2 \chi(r)}{dr^2} + \left\{(E-V)^2 - (m+S)^2 - \frac{\kappa(\kappa-1)}{r^2}\right\} \chi(r) = \\ = \left\{\frac{3}{4}\frac{(V'-S')^2}{(E-V-m-S)^2} + \frac{V'' - S'' + \frac{2\kappa}{r}(V'-S')}{2(E-V-m-S)}\right\} \chi(r) \tag{45}$$

Accordingly

$$G(0) = 0 \tag{46}$$

$$\lim_{r \to 0} G(r) \approx r^{N_1}; \quad N_1 = 0, 1, 2, \ldots \tag{47}$$

These equations are also known in scientific literature [18-20]. But here we want to stress that *they acquire their status only after fulfillment of boundary conditions (39,46) by functions behaving at the origin as in (40),(47). Only in this case the solutions of radial equations (43),(45) satisfy to the primary total 3-dimensional Dirac equation [21].* We see also that the conditions (39) and (46) are in accordance with the time-independence of the norm.

Equations (43) and (45) have a similar structure: their left-hand sides coincide to the Klein-Gordon operation, while the right-hand-sides include corrections of spin-orbit interaction.

## 5. Application of radial equations

As a first application let us consider regular potentials with positive degrees like $r^m$, $m > 0$ both for $V$ and $S$, which are regular everywhere except infinity and mainly are used in quark confinement problems. For such potentials infinite distances are relevant, therefore asymptotics at infinity is interesting. It follows from equation (43) as $r \to \infty$, that

$$\frac{d^2 \varphi}{dr^2} + (V^2 - S^2)\varphi = 0 \tag{48}$$

We see that only vector potential does not provide a decreasing behavior of wave function at infinity (Klein paradox), whereas the scalar potential or their mixture with scalar domination does. Of course, it is a well-known fact [22]. The following question arises naturally: *what kind of potentials can we use in the reduced equations (43) and (45)?*

The behavior at the origin is able to answer this question. Consider potentials like (30). The case $\alpha > 1$ is not interesting, because it leads to the falling onto the center. For potentials (30) the right hand sides of Eq.(43) and (45) are dominated by $r^{-2}$ behaving terms, and in the left hand side only centrifugal term $\kappa(\kappa \pm 1)/r^2$ dominates in this case, both equations (43,45) have the following leading asymptotic

$$\varphi'' - \left\{\left(\frac{1-\alpha}{2}+\kappa\right)^2 - \frac{1}{4}\right\}\frac{\varphi(r)}{r^2} = 0; \quad \chi'' - \left\{\left(\frac{1-\alpha}{2}-\kappa\right)^2 - \frac{1}{4}\right\}\frac{\chi(r)}{r^2} = 0 \quad (49)$$

It follows from these equations and from (42-44), that the reduced wave functions behave like

$$F(r) \underset{r\to 0}{\approx} \sqrt{V_0} r^{-\frac{\alpha}{2}}\left\{Ar^{1/2+P_1} + Br^{1/2-P_1}\right\}; \quad G(r) \underset{r\to 0}{\approx} \sqrt{V_0} r^{-\frac{\alpha}{2}}\left\{Ar^{1/2+P_2} + Br^{1/2-P_{21}}\right\} \quad (50)$$

where

$$P_1 = \left|\frac{1-\alpha}{2}+\kappa\right|; \quad P_2 = \left|\frac{1-\alpha}{2}-\kappa\right| \quad (51)$$

We see that the degree in (50) never be an integer. Therefore, (40) and (47) conditions are not fulfilled and so single equations (43), (45) are not permissible. One can consider only the case $\alpha = 0$, which corresponds to constant walls or logarithmic potential, or free motion.

In the case of Coulomb potential (20) the left hand-side of eq.(43) participates also to the leading behavior, therefore wave functions have the distinguished behavior at the origin

$$F(r) \underset{r\to 0}{\approx} \sqrt{V_0} r^{-1/2}\left\{d_1 r^{1/2+P_3} + d_2 r^{1/2-P_3}\right\}; \quad P_3 = \sqrt{k^2 + 1/4 - V_0^2} \quad (52)$$

We see that the degree here is also fractional, which is not desirable, because from the condition (40) we have $1/2 \pm P_3 = N$ and it leads to the strange quantization of $V_0$ constant

$$V_0^2 = \kappa^2 - N(N-1); \quad N = 1,2,3... \quad (53)$$

It follows that, in this case, there are no solutions except for ''quantized'' $V_0$ and this is senseless.

One remark is in order for Coulomb potential. According to (52), its asymptotic depends on module $|\kappa|$, which probably means that the energy spectrum for this potential depends only on $|\kappa| = j + 1/2$. This conclusion is in accordance with the explicit solution of the Coulomb problem – the Sommerfeld formula [23], with all other consequences, such as absence of the Lamb shift, Witten's N=2 superalgebra and etc.[24].

## 6. Conclusions

In this article, we have studied the Dirac radial equation. Behavior of radial function at the origin is assigned in the general framework of Quantum mechanics. It has been established that the so-called additional solution's behavior at the origin contradicts to the normalization property, because besides vanishing of reduced radial function some definite turning to zero must be satisfied according to distribution theory and therefore the Dirac equation does not allow for "hydrino"-like solutions. Growing of singularity at the origin in radial Dirac equation is explained in detail and the role of spin in this phenomenon is cleared up.

We have shown also that in the process of passing to the second order single equation the delta-function containing term appears, avoidance of which and derivation of commonly known equation is related to very strict mathematical constraint on radial functions – they must be infinitely smooth functions with derivatives of infinite times, (See Eqs.(40) or (47)).

This restriction is severe really. As a consequence, the second order reduced radial equations (43) and (45) are not valid practically for all potentials, besides problems of free particle and potential walls. Logarithmic potential is also available, but for it behavior of up and down components has different asymptotics at the origin.

In other words, we can freely work only with radial equations (35) or (35a), i.e. before the exclusion of the first derivative terms.

It must be noted that the appearance of delta-function in the Coulomb problem was mentioned earlier in the paper of R. Armstrong and E. Power [25]. But this idea, unfortunately, had not been responded during large time. The strict mathematical substantiation, as a singularity of the radial Laplacian, this fact acquires in [10,11]. Interesting enough, that this fact has rather general character, because it does not depend on particular potential – is it regular or singular. Only character of turning to zero depends on potential. This problem takes place also in Laplace operators on three- and more- dimensions [10].

Many problems of physics are used to considered in less (n<3) dimensions as well, mainly in condensed matter physics. We will see below that in such cases delta-like singularity does not appear. Therefore it is natural and highly desirable to find some adequate physical requirements to avoid appearance of additional solutions.

Because of current interest to this problem we have included below the Appendix B for case of 2-dimensional Dirac equation in Coulomb field.

Anzor Khelashvili, Teimuraz Nadareishvili

**Appendix A: Spin effects and nonexistence of "Hydrino" states**

In this appendix we extract the spin role more explicitly in problem under consideration. Let us use the equation (35), rewriting it in the form (for simplicity, we consider again only vector case)

$$f'' + \frac{2}{r}f' + \left[(E-V)^2 - m^2 - \frac{\kappa(\kappa+1)}{r^2}\right]f = \frac{V'}{E+m-V}f' + \frac{V'}{E+m-V}\frac{1+\kappa}{r}f \quad (A.1)$$

It is evident that the left-hand side gives the Klein-Gordon operation, while the right hand side accounts the spin corrections. Consider potentials like

$$\lim_{r\to 0} rV(r) = -V_0 ; \quad (A.2)$$

and substitute behavior of radial function in the form $f \approx r^\lambda$. We obtain the following characteristic equation

$$\lambda(\lambda-1) + 2\lambda + \left[V_0^2 - \kappa(\kappa+1)\right] = -(\lambda+1+\kappa) \quad (A.3)$$

If we ignore here the right-hand side (i.e. spin influence), it follows

$$\lambda^2 + \lambda + \left[V_0^2 - \kappa(\kappa+1)\right] = 0 \quad (A.4)$$

from which

$$\lambda = -\frac{1}{2} + P; \quad P = \sqrt{(\kappa+1/2)^2 - V_0^2} \quad (A.5)$$

Therefore, if we do not take spin into account, we derive the following behavior at the origin

$$\lim_{r\to 0} f = a_{st} r^{-1/2+P} + a_{add} r^{-1/2-P} \equiv f_{st} + f_{add} \quad (A.6)$$

and we must include the additional state when $0 < P < 1/2$ and $\kappa(\kappa+1) < V_0^2$. This situation appears in the Schrodinger and Klein-Gordon equations.

On the other hand, if we account spin (i.e. the right-hand side in Eq.(A.3)), we obtain from (A.2) the equation

$$\lambda^2 + 2\lambda + \left[V_0^2 - \kappa^2 + 1\right] = 0 \quad (A.7)$$

We see that in front of $\lambda$ the extra factor $2$ appears, because of it the singularity at the origin rises and the behavior (25) appears, therefore there are no additional states, that will be consistent to the general principles of quantum mechanics.

Dirac's radial equations and the Additional Solutions

**Appendix B. The Dirac equation in two dimensions**

Two dimensional Dirac equation for studding existence of additional states (hydrino) and for self-adjoint extension procedure was considered in many papers [3, 4, 26, 27]. Below we'll proceed to the work of Dombey [4]. Let's remark in parallel that the conclusion of this work about non-existence of hydrino state in the Klein-Gordon equation is premature, because in the limit $\alpha \to 0$ the Coulomb interaction is put out and the problem reduces to free motion, but not in a Coulomb field. If we take the approximation $\alpha^2 \ll \alpha$, as is done in this article, then in the first order of $\alpha$ there remains a tightly bound state with $E_{0A} = m\alpha$, i.e. the additional solution which has no non-relativistic analogue. Such kind of phenomenon is known still from the Bethe-Salpeter equation, so-called, the Wick-Cutkosky model [28,29] and the repetition of tightly bound state in the Klein-Gordon equation is not surprising.

As regards of Dirac 2-dimensional equation in radial variables it has the form [4]

$$f' - \frac{j-1/2}{r}f + (E+m-V)g = 0$$
$$g' + \frac{j+1/2}{r}g - (E-m-V)f = 0$$
(B.1)

Now we'll follow the same strategy as for 3-dimensional case, described above, namely determine $g$ from the second equation and substitute into the first one. It follows

$$f'' + \frac{1}{r}f' + \frac{V'f'}{E+m-V} - \frac{j-1/2}{r}\frac{V'}{E+m-V}f + \left[(E-V)^2 - m^2 - \frac{(j-1/2)^2}{r^2}\right]f = 0$$
(B.2)

At the same manner we derive the equation for $g$:

$$g'' + \frac{1}{2}g' + \frac{V'g'}{E-m-V} + \frac{j+1/2}{r}\frac{V'g}{E-m-V} + \left[(E-V)^2 - m^2 - \frac{(j-1/2)^2}{r^2}\right]g = 0 \quad (B.3)$$

One can exclude the first derivatives in the first two terms by substitutions

$$f = \frac{A}{\sqrt{r}}; \qquad g = \frac{B}{\sqrt{r}}$$
(B.4)

This is analogous of transformations (36) of the main text. But now

$$\left(\frac{d^2}{dr^2} + \frac{1}{r}\frac{d}{dr}\right)\left(\frac{1}{\sqrt{r}}\right) = \frac{1}{4}r^{-5/2}$$
(B.5)

which is not a delta function in difference of 3-dimensional case. Therefore wave functions are not restricted whatever and we obtain equations free from delta function

Anzor Khelashvili, Teimuraz Nadareishvili

$$\left[\frac{d^2}{dr^2} - \frac{j(j-1)}{r^2}\right]A(r) + \frac{V'}{E+m-V}A' - \frac{j}{r}\frac{V'}{E+m-V}A + \left[(E-V)^2 - m^2\right]A(r) = 0 \quad \text{(B.6)}$$

One can exclude the first derivative completely by substitution

$$A(r) = \sqrt{E+m-V}\,\varphi(r) \quad \text{(B.7)}$$

Then we derive the equation similar to Eq. (43):

$$\frac{d^2\varphi(r)}{dr^2} + \left\{(E-V)^2 - m^2 - \frac{j(j-1)}{r^2}\right\}\varphi(r) = \left\{\frac{3}{4}\frac{(V')^2}{(E-V+m)^2} + \frac{V'' + \frac{2j}{r}V'}{2(E-V+m)}\right\}\varphi(r) \quad \text{(B.8)}$$

Quite analogously can be derived the final equation for $g(r)$. The difference with 3-dimensional equations is that now the equation is valid without some restriction of wave functions and, consequently, the solutions of (B.5) satisfy to the primary total 2-dimensional Dirac equation.

The non-relativistic limit in Eq. (B.8) may be performed as follows: first of all we neglect the right-hand side – contribution from spin-orbital coupling, then the Klein-Gordon equation remains, in which one makes the changing

$$(E-V)^2 - m^2 = (E-V-m)(E-V+m) \approx$$
$$\approx (\varepsilon - V)(2m - V) \approx 2m(\varepsilon - V)$$

At the last step we have ignored $V \ll 2m$ and $\varepsilon$ is a binding energy. After this we are faced with the Schrodinger equation. Therefore our equation has correct non-relativistic limit.

Further steps for additional solutions are to check several physical requirements, as orthogonality and etc. This is easily done by using the first form of equation (B.1) for the Coulomb potential $V = -\frac{\alpha}{r}$, and using the substitutions (B.2). It follows the equations

$$\frac{dA}{dr} - j\frac{A}{r} + \left(E + \frac{\alpha}{r} + m\right)B = 0$$
$$\frac{dB}{dr} + j\frac{B}{r} + \left(m - E - \frac{\alpha}{r}\right)A = 0 \quad \text{(B.9)}$$

Let us now write down these equations for arbitrary two levels $E_1$ and $E_2$, then multiply first pair of equations on $A_1$ and $A_2$, the second pair – on $B_1$ and $B_2$, correspondingly and consider their sum and difference. In result we derive the relation:

$$(E_2 - E_1)(A_1 A_2 + B_1 B_2) = A_2 B_1' - A_1 B_2' + B_1 A_2' - B_2 A_1' \quad \text{(B.10)}$$

which after partial integration on the right-hand side gives

$$(E_2 - E_1)\int_0^\infty (A_1 A_2 + B_1 B_2)\,dr = \lim_{r \to 0}(A_2 B_1 - A_1 B_2) \qquad (B.11)$$

where we have taken into account that the bound state wave function goes to zero at infinity.

Finally, let us use this relation for solutions in the Coulomb field from the article of Dombey [4], (see, discussion after (23)): "As $r \to 0, f, g \approx r^t = r^{-1/2+\sqrt{1/4-\alpha^2}}$ (for solution, which is square integrable down to the origin even they diverge"). We see that this type of solutions are orthogonal as well.

"The anomalous solutions arise from the other root, where $t = -1/2 - \sqrt{1/4-\alpha^2}$. They too are square integrable down to the origin and this solution has no non-relativistic analogue and like the hydrino corresponds to a state which is tightly bound for small $\alpha$" [4].

It is seen that nevertheless these anomalous solution do not satisfy the orthogonality property (B.11). Indeed, the solution gives

$$A \approx a r^{-\sqrt{1/4-\alpha^2}}, \qquad B \approx b r^{-\sqrt{1/4-\alpha^2}} \qquad (B.12)$$

Therefore the right-hand side of orthogonality relation (B.7) becomes

$$(E_2 - E_1)\int_0^\infty (A_1 A_2 + B_1 B_2)\,dr = \lim_{r \to 0} r^{-2\sqrt{1/4-\alpha^2}}(a_2 b_1 - a_1 b_2) \qquad (B.13)$$

As coefficients here are arbitrary, we see that this expression diverges and therefore, the hydrino states do not satisfy to orthogonality relation, moreover for this solution the differential probability in the slice $(r, r+dr)$ is also divergent at the origin and the norm of these solution is not time-independent. These arguments are sufficient for their ignorance [11].

In conclusion, in two-dimensions, as well as in three-dimensions anomalous (hydrino) solutions do not satisfy to fundamental physical principles and their existence is not possible.

Finally, let us remember that some authors consider nucleus not as point particle but as charge extending over an arbitrarily small but finite radius and eliminating additional anomalous solutions. This approach is not new, of course. It was considered still in the teaching book Landau and Lifshitz [14] in context of ordinary quantum mechanics for avoiding the extra solution in $r^{-2}$-like potential problem. However, by our opinion this approach is equivalent to considering of a different Hamiltonian and lead to discarding of possible bound state, which arises after the self-adjoint extension procedure ([9],Ch. 3).

Anzor Khelashvili, Teimuraz Nadareishvili

**Acknowledgments**

This work was supported by Shota Rustaveli National Science Foundation (SRNSF) [grant number № DI-2016-26, Project Title: "Three-particle problem in a box and in the continuum''']. We are also indebted to anonymous referee for get our attention to the two-dimensional Dirac problem.

**References**

1. J. Naudts. arXiv: 0507193 [Physics] (2005).
2. P.Giri. arXiv: 080833.09 [cond-mat.mtrl-sci] (2008).
3. A.Rathke. New Journal of Physics **7,** 127(2005).
4. N.Dombey. Phys. Lett. A **360,** 62 (2006).
5. A. De Castro. Phys. Lett. A **369,**380 (2007).
6. J.Va'vra J. Astronomy and Astrophysics manuscripts.ESO (2013).
7. A. Mauleberg and K.Sinha. J. Cond.Matter Nuclear Sci. **4,** 241(2011).
8. A. Mauleberg. J. Cond. Matter Nucl. Sci.**10** 15(2013).
9. T.Nadareishvili and A.Khelashvili A. arXiv: 0903.0234[math-ph] (2009).
10.Y. Cantelaube and L.Khelif. J.Math.Phys. **51** 053518(2010).
11. A.Khelashvili and T.Nadareishvili. Am. J. Phys. **79** 668 (2011); arXiv:1001.3285
12. A.Khelashvili and T. Nadareishvili.Phys. of Particles and Nuclear Lett. **12** 11(2015).
13. J. Maly and J. Va'vra.Fusion Technology **24**.No2 (1993).
14. L. Landau and E. Lifshitz. *Quantum Electrodynamics* (Pergamon press. 1982).
15.W.Pauli.*Die allgemeinen Prinzipen der Wellenmechanik* (In Handbuch der Fizik,Bd.5.Vol 1.Berlin:Auf, 1958).
16. A. Khelashvili and T. Nadareishvili. European J.Phys **35** 065026 (2014).
17. I.Gelfand and G.Shilov. *Generalized functions (*Moscow,Nauka) (in Russian), 1958).
18. N. Poliatzky.Phys.Rev.Lett. **70** 2507(1993)
19. A. Soylu, O. Bayrak O and I.Boztosun I. J.Phys.A **41** 065308 (2008).
20.Ying Xu, Su He and Chun-Sheng Jia. J.Phys.A **41** 255302(2008)
21.P A M. Dirac.*The Principles of Quantum Mechanics* (Oxford:At the Clarendon Press,1958).
22.D. Bjorken and D. Drell. *Relativistic Quantum Mechanics* (McGraw-Hill Book Company,1964).
23. A.J.W. Sommerfeld. Ann. physik **51**,125 (1916).
24.T.Khachidze and A.Khelashvili.*Dynamical symmetry of the Kepler-Coulomb problem in classical and Quantum mechanics*. (Nova Science Publishers,Inc. New York, 2008).
25.B.H.Armstrong and E.A.Power.Am.J.Phys**. 31** 262 (1963).
26. F. M. Andrade, E.O. Silva, T. Prudêncio and C. Filgueiras. J. Phys. G: Nucl. Part. Phys. **40** 075007(2013).
27. H. Falomir and P.A.G. Pisani. J.Phys.A**34** 4143 (2001).

Dirac's radial equations and the Additional Solutions

28. G.C.Wick.Phys.Rev. **96** 1124 (1954).
29. R.E.Cutkosky. Phys. Rev. **96** 1135 (1954).